\documentclass[12pt]{ussci}
\pdfoutput=1
\usepackage{xmpincl}
\includexmp{CC_Attribution_4.0_International}

\hypersetup{
    pdftitle={High-Pressure Autoignition of Binary Blends of Methanol and Dimethyl Ether},%
    pdfauthor={Bryan W. Weber},%
}

\usepackage{enumitem}
\setlist{noitemsep}

\usepackage[version=4]{mhchem}
\usepackage{siunitx}
\usepackage{graphicx}
\usepackage{pgf}
\usepgflibrary{fpu}

\usepackage[tableposition=top]{caption}
\usepackage{booktabs}
\usepackage{mathtools}
\usepackage{microtype}

\usepackage[capitalize]{cleveref}
\newcommand\papertopic{Reaction Kinetics}
\DeclareSIUnit\torr{torr}

\title{ High-Pressure Autoignition of Binary Blends of Methanol and Dimethyl Ether }

\author[1,2]{Hongfu Wang}
\author[2*]{Bryan W.\ Weber}
\author[2]{Ruozhou Fang}
\author[2]{Chih-Jen Sung}
\affil[1]{School of Mechanical and Electrical Engineering, Nanchang University, Jiangxi Province, P.R. China}
\affil[2]{Department of Mechanical Engineering, University of Connecticut, Storrs,
CT, USA}
\affil[*]{Corresponding Author: \email{bryan.weber@uconn.edu}}

\begin{document}
\maketitle

\begin{abstract} 

    Reactivity Controlled Compression Ignition (RCCI) is a new advanced engine
    concept that uses a dual fuel mode of operation to achieve significant
    improvements in fuel economy and emissions output. The fuels that are
    typically used in this mode include a low- and a high-reactivity fuel in
    varying proportions to control ignition timing. As such, understanding the
    interaction effects during autoignition of binary fuel blends is critical
    to optimizing these RCCI engines. In this work, we measure the autoignition
    delays of binary blends of dimethyl ether (\ce{C2H6O}, DME) and methanol
    (\ce{CH4O}, MeOH) in a rapid compression machine. In these experiments,
    dimethyl ether and methanol function as the high- and low-reactivity fuels,
    respectively. We considered five fuel blends at varying blending ratios (by
    mole), including \SI{100}{\percent} DME-\SI{0}{\percent} MeOH,
    \SI{75}{\percent} DME-\SI{25}{\percent} MeOH, \SI{50}{\percent}
    DME-\SI{50}{\percent} MeOH, \SI{25}{\percent} DME-\SI{75}{\percent} MeOH,
    and \SI{0}{\percent} DME-\SI{100}{\percent} MeOH. Experiments are conducted
    at an engine-relevant pressure of \SI{30}{\bar}, for the stoichiometric
    equivalence ratio. In addition, the experimental results are compared with
    simulations using a chemical kinetic model for DME/MeOH combustion generated
    by merging independent, well-validated models for DME and MeOH.

\end{abstract}


\section{Introduction}\label{introduction}

To reduce the environmental impact of combustion, future combustion processes
must feature substantially reduced pollutant emissions while maintaining high
efficiency. A promising concept in this respect is low-temperature combustion
(LTC). As an outstanding representative of LTC technologies, the dual-fuel
Reactivity Controlled Compression Ignition (RCCI) operation has great potential
in terms of combustion controllability. The general principle of dual-fuel RCCI
combustion requires two fuels with different reactivities, that is using a high
reactivity fuel (such as diesel or dimethyl ether (DME)) to trigger the ignition
and combustion of low-reactivity fuels (such as gasoline, methanol, ethanol, or
butanol).

DME is considered an efficient alternative fuel for use in diesel engines
because it has excellent autoignition characteristics. The low boiling point
(\SI{248}{\K}), low critical point (\SI{400}{\K}), and high cetane number (\(>
55\)) of DME~\autocite{Arcoumanis2008,Teng2001} make it well suited for
compression ignition engines. In addition, the high oxygen content of DME
(\SI{34.8}{\percent} by mass) together with the absence of C-C bonds
contributes to ultra-low soot formation during DME
combustion~\autocite{Arcoumanis2008}.

The promoting effect of DME blending on fuels with poor autoignition qualities
such as methane and propane is a promising feature with respect to the RCCI
concept. Therefore, some research has been conducted to reveal the promoting
potential of DME. Several researchers~\autocite{Burke2015a,Tang2012a,Chen2007a}
have studied the ignition delays of DME/methane blends in shock tubes. The
results have shown that DME has a strong promoting effect on the autoignition of
methane, even when the concentration of methane is much higher than that of DME.
Further, \textcite{Dames2016} showed the promoting effect of DME in DME/propane
blends using a rapid compression machine. The results of the work of
\textcite{Dames2016} showed that propane combustion is promoted due to the large
amount of radicals produced by low-temperature DME oxidation. These studies
indicate the potential of DME as a combustion promoter together with fuels with
poor auto-ignition qualities.

For dual-fuel RCCI operation, a lower reactivity fuel should be studied in
combination with the high reactivity fuel. Methanol (MeOH) is well known as a
widely used alcoholic alternative fuel, but when operated in a single fuel mode
it tends to have poor autoignition quality~\autocite{Siebers1987}. Given the
great potential of DME as a combustion promoter, we expect that DME can
significantly improve the auto-ignition quality of methanol and the combination
will be effective for dual-fuel RCCI operation.

In this work, we explore the autoignition characteristics of DME/MeOH binary
blends. A set of ignition delay time data for DME/MeOH blends at different
blending ratios over a wide range of temperature at engine relevant pressure
conditions and the stoichiometric equivalence ratio is obtained in a rapid
compression machine (RCM). In addition, we compile a chemical kinetic model for
DME/MeOH combustion by merging independent models for the fuels. Simulations
utilizing this model are compared to experimental results and good agreement is
observed over the range of the experiments.

\section{Experimental Methods}\label{sec:experimental-methods}

\begin{figure}[htb]
    \centering
    \resizebox{0.6\textwidth}{!}{\input{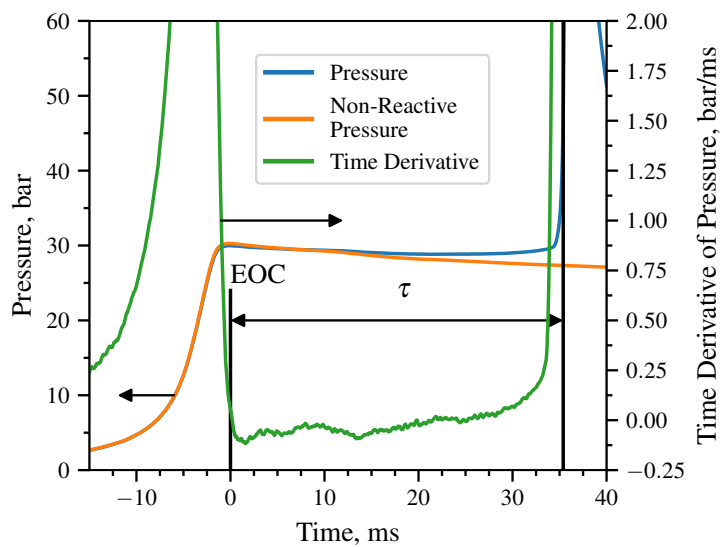}}
    \caption{Definition of the ignition delay used in this work. The
    experiment in this figure was conducted for a \SI{25}{\percent} DME blend
    with \(P_0=\SI{1.1528}{\bar}\), \(T_0=\SI{338}{\K}\),
    \(P_C=\SI{29.98}{\bar}\), \(T_C=\SI{748}{\K}\), and
    \(\tau=\SI{35.39\pm1.93}{\ms}\).}
    \label{fig:ign-delay-def}
\end{figure}

The RCM used in this study is a single piston arrangement and is pneumatically
driven and hydraulically stopped. The device has been described in detail
previously~\autocite{Mittal2007a} and will be described here briefly for
reference. The end of compression (EOC) temperature and pressure (\(T_C\) and
\(P_C\) respectively), are independently changed by varying the overall
compression ratio, initial pressure, and initial temperature of the experiments.
The primary diagnostic on the RCM is the in-cylinder pressure. The pressure data
is processed by a Python package called UConnRCMPy~\autocite{uconnrcmpy}, which
calculates \(P_C\), \(T_C\), and the ignition delay(s). The definition of the
ignition delay is shown in \cref{fig:ign-delay-def}. The time of the EOC is
defined as the maximum of the pressure trace prior to the start of ignition and
the ignition delay is defined as the time from the EOC until local maxima in
the first time derivative of the pressure.

In addition to the reactive experiments, non-reactive experiments are conducted
to determine the influence of machine-specific behavior on the experimental
conditions and permit the calculation of the EOC temperature via the isentropic
relations between pressure and temperature~\autocite{Lee1998}. The EOC
temperature is calculated by the procedure described in
\cref{sec:rcm-modeling}.

The RCM is equipped with heaters to control the initial temperature of the
mixture. After filling in the components to the mixing tanks, the heaters are
switched on and the system is allowed \SI{1.5}{\hour} to come to steady state.
The mixing tanks are also equipped with magnetic stir bars so the reactants are
well mixed for the duration of the experiments.

The mixtures considered in this study are shown in \cref{tab:mixtures}. The
``\si{\percent} DME'' and ``\si{\percent} MeOH'' columns indicate the molar
percent of each component in the fuel blend. Mixtures are prepared in stainless
steel mixing tanks. The proportions of reactants in the mixture are determined
by specifying the absolute mass of the methanol in the mixture (if present), the
equivalence ratio, the oxidizer composition (in this study, \ce{O2} and \ce{N2}
in the ratio of $1:3.76$ are used throughout), and the molar ratio of DME/MeOH
in the fuel blend. Since MeOH is a liquid at room temperature and pressure, it
is injected into the mixing tank through a septum. Proportions of DME, \ce{O2},
and \ce{N2} are added manometrically at room temperature.

\begin{table}[htb]
    \centering
    \caption{Mixtures considered in this work}
    \begin{tabular}{SScccc}
        \toprule
        && \multicolumn{4}{c}{Mole Fraction (purity)} \\
        \cmidrule{3-6}
        {\si{\percent} DME}& {\si{\percent} MeOH} & DME (\SI{99.7}{\percent}) & MeOH (\SI{100.00}{\percent}) & \ce{O2} (\SI{99.994}{\percent}) & \ce{N2} (\SI{99.999}{\percent})  \\
        \midrule
        100 & 0 & 0.0654 & 0.0000 & 0.1963 & 0.7383 \\
        75 & 25 & 0.0556 & 0.0185 & 0.1945 & 0.7314 \\
        50 & 50 & 0.0427 & 0.0427 & 0.1921 & 0.7225 \\
        25 & 75 & 0.0252 & 0.0756 & 0.1889 & 0.7103 \\
        0 & 100 & 0.0000 & 0.1229 & 0.1843 & 0.6928 \\
        \bottomrule
    \end{tabular}
    \label{tab:mixtures}
\end{table}

\section{Computational Methods}\label{sec:computational-methods}

To the best of our knowledge, there are no chemical kinetic models for the
combustion of binary blends of DME and MeOH available in the literature.
Therefore, we compile a kinetic model in this work by combining two independent
models. The kinetics for DME are taken from the work of \textcite{Burke2015a}
while the kinetics for MeOH are taken from the work of \textcite{Burke2016}.

To combine the two models, duplicate reactions and species were taken from the
\textcite{Burke2015a} model; however, the models were produced by the same
research group approximately one year apart, so we do not expect many
differences in the common chemistry. In the work of \textcite{Dames2016}, it was
found that for combined models of high-reactivity fuels such as DME and
low-reactivity fuels such as propane, cross-reactions between the fuels do not
strongly affect the ignition delay and the fuels instead interact through
radical species such as \ce{OH}. Therefore, we do not consider any
cross-reactions between the high- and low-reactivity fuels in this study (DME
and MeOH, respectively).

\subsection{RCM Modeling}\label{sec:rcm-modeling}

All of the simulations in this work use the Python interface for Cantera 2.3.0
\autocite{cantera}. Two types of simulations are considered. The first is used
to calculate \(T_C\). Detailed descriptions of the use of Cantera for these
simulations can be found in the work of \textcite{Weber2016a} and
\textcite{Dames2016}; a brief overview is given here. As mentioned in
\cref{sec:experimental-methods}, non-reactive experiments are conducted to
characterize the machine-specific effects on the experimental conditions in the
RCM. This pressure trace is used to compute a volume trace by assuming that the
reactants undergo a reversible, adiabatic, constant composition (i.e.,
isentropic) compression during the compression stroke and an isentropic
expansion after the EOC. The volume trace is applied to a non-reactive
simulation conducted in an \verb|IdealGasReactor| in Cantera \autocite{cantera}
and the temperature at the end of compression is reported as \(T_C\).

The second type of simulation uses a constant-volume, adiabatic reactor. This
method does not consider the effect of the compression stroke and
post-compression heat loss present in the experiments and the initial conditions
in the simulation are set equal to the EOC conditions in the experiment. The
ignition delay is defined as the time required for the simulated temperature to
increase by \SI{400}{\K} over the initial temperature in the simulation.

\section{Results and Discussion}\label{sec:results-and-discussion}

Ignition delay results for the mixtures listed in \cref{tab:mixtures} are shown
in \cref{fig:ign-delays} for the stoichiometric equivalence ratio and \(P_C =
\SI[number-unit-product={\ }]{30}{\bar}\). In the following, we use the
shorthand of specifying the molar percent of DME in the fuel blend to indicate
the blending condition.

It can be seen in \cref{fig:ign-delays} that the \SI{100}{\percent} DME case
(\SI{0}{\percent} MeOH) is the most reactive while the \SI{0}{\percent} DME case
(\SI{100}{\percent} MeOH) is the least reactive. Interestingly, the change in
reactivity as MeOH is added to DME appears to be non-linear with respect to the
molar percent of MeOH added. In other words, the change in ignition delay at a
fixed \(T_C\) is smaller going from \SI{100}{\percent} DME to \SI{50}{\percent}
DME than going from \SI{50}{\percent} DME to \SI{0}{\percent} DME.

This is also demonstrated by \cref{fig:temp-comp}, which shows the \(T_C\)
values for ignition delays near \SI{20}{\ms} at the range of mixtures considered
in this study. As the \si{\percent} DME decreases in the blend, the temperature
required to achieve the same ignition delay increases. However, the temperature
increase from \SI{100}{\percent} to \SI{50}{\percent} DME is much smaller than
the increase from \SI{50}{\percent} to \SI{0}{\percent} DME.

\begin{figure}[htb]
    \begin{minipage}[t]{0.48\textwidth}
        \centering
        \resizebox{\linewidth}{!}{\input{figures/simulation-comparison.pgf}}
        \caption{Ignition delays of blends of DME and MeOH as a function of
        inverse temperature, for an equivalence ratio of \(\phi = 1.0\) and
        \(P_C = \SI[number-unit-product={\ }]{30}{\bar}\). Constant volume,
        adiabatic simulations are shown as the solid lines.}
        \label{fig:ign-delays}
    \end{minipage}\hfill%
    \begin{minipage}[t]{0.48\textwidth}
        \centering
        \resizebox{\linewidth}{!}{\input{figures/temperature-comparison.pgf}}
        \caption{\(T_C\) values for ignition delays near \SI{20}{\ms} at the
        range of blends considered in this study}
        \label{fig:temp-comp}
    \end{minipage}\hfill%
\end{figure}

Also shown on \cref{fig:ign-delays} are constant volume, adiabatic simulations
computed according to the procedure laid out in \cref{sec:rcm-modeling}. In
general, the agreement between the model and the experiments is quite good over
the entire range of the experiments. It can be seen in \cref{fig:ign-delays}
that at low temperatures for a given mixture composition the ignition delay
tends to be under-predicted, while at the higher temperatures the ignition delay
is over-predicted.

As discussed by \textcite{Mittal2008}, this is likely due in part to the
modeling procedure used in this work. In general, we expect constant volume
simulations to have shorter ignition delays than the experiments for long
ignition delays because they do not include the effect of post-compression heat
loss; conducting simulations that include the post-compression heat loss are
very likely to improve agreement in this region. Furthermore, for short ignition
delays, we expect constant volume simulations to over-predict the experimental
ignition delay because they do not include the effect of radical pool buildup
during the compression stroke. Therefore, conducting simulations that include
the compression stroke are very likely to improve the agreement for short
ignition delays.

\section{Conclusions}\label{sec:conclusions}

In this study, we have measured ignition delays for binary blends of dimethyl
ether and methanol for engine-relevant pressure, temperature, and equivalence
ratio conditions using a heated rapid compression machine. The ignition delay
results show that pure DME is more reactive than pure MeOH, and that the
increase in ignition delay as DME is replaced by MeOH is non-linear as a
function of the blending fraction. The ignition delays are also compared to a
chemical kinetic model compiled by combining independent models for the two
fuels. This model does not consider cross reactions between DME and MeOH.
Nonetheless, the model gives quite good agreement with the data, supporting the
hypothesis that the fuels do not interact via cross reactions but instead
through common radicals such as \ce{OH}. In addition, this further demonstrates
that models for low-reactivity fuels such as methanol and high-reactivity fuels
such as DME can be constructed by simple concatenation and deduplication of
their respective independent models.

\section{Acknowledgements}\label{acknowledgements}

This work was supported by the National Science Foundation under Grant No.
CBET-1402231.

\printbibliography

\end{document}